\documentclass[preprint,aps,floatfix,nofootinbib]{revtex4}
\usepackage{graphicx}
\usepackage{bbm}
\newcommand{\fr}{\frac}      
\newcommand{\bea}{\begin{eqnarray}}
\newcommand{\eea}{\end{eqnarray}}
\begin{document}
\preprint{\vbox{\hbox{JLAB-THY-03-25} \hbox{DUKE-TH-03-238} \hbox{ECT$^*$-03-08} }}
\vspace{0.5cm}
\title{\phantom{x}
\vspace{0.5cm} Analysis of the $[{\mathbf{56}},2^+]$  Baryon Masses  in
the  ${\mathbf 1/N_c}$ Expansion}

\author{
J. L. Goity $^{a,b}$ \thanks{e-mail: goity@jlab.org}, \ C.
Schat $^{c}\ $ \thanks{e-mail: schat@phy.duke.edu}, \ N. N.
Scoccola $^{d,e,f}$ \footnote[2]{Fellow of CONICET, Argentina.}
\thanks{e-mail: scoccola@tandar.cnea.gov.ar}}

\affiliation{
$^a$ Department of Physics, Hampton University, Hampton, VA 23668, USA. \\
$^b$  Thomas Jefferson National Accelerator Facility, Newport News, VA 23606, USA. \\
$^c$ Department of Physics, Duke University, Durham, NC  27708, USA.  \\
$^d$ Physics Department, Comisi\'on Nacional de Energ\'{\i}a At\'omica,
     (1429) Buenos Aires, Argentina.\\
$^e$ Universidad Favaloro, Sol{\'\i}s 453, (1078) Buenos Aires, Argentina.\\
$^f$ ECT$^*$, Villa Tambosi, I-38050 Villazzano (Trento) Italy.}
\date{\today}
\begin{abstract}
The mass spectrum of the  positive parity $[{\mathbf{56}},2^+]$  baryons is studied in the
$1/N_c$ expansion up to and including  ${\cal{O}}(1/N_c)$ effects with  $SU(3)$
symmetry breaking implemented to first order. A total of eighteen mass relations
result, several of which are tested with the available data. The breaking of spin-flavor symmetry is dominated by 
the hyperfine interactions, while spin-orbit effects are found to be small. 
\end{abstract}
\maketitle

In the mass range from 1600 to 2100 MeV there exists a set of positive parity baryons
which might be assigned to an irreducible representation  $[{\mathbf{56}},2^+]$
of  $SU(6) \otimes O(3) $, where $SU(6)$ is the spin-flavor group and $O(3)$
classifies the orbital excitations.
Among the candidate states in  that set, all non-strange states are known as well as
seven strangeness ${\cal S} = -1$ states. Some of the strange states are, however, established
with low  certainty (two stars or less in the Particle Data Listings \cite{PDG02}). In
this letter the available empirical information is used to implement an analysis of the masses based on
the $1/N_c$ expansion of QCD \cite{tHo74,Wit79}, an approach that  has turned out to be very successful in baryon  phenomenology.  

The $1/N_c$ expansion has been applied to  the ground state baryons \cite{DM93,DJM94,DJM95,CGO94,Jenk1,JL95,DDJM96}, 
and to excited baryons, where the masses and decays of the negative parity spin-flavor $\mathbf{70}$-plet 
\cite{Goi97,PY,CCGL,CaCa98,SGS} and the positive parity Roper $\mathbf{56}$-plet \cite{CC00} have been analyzed.
Two frameworks have been used in implementing the $1/N_c$ expansion for baryons.
One framework is based on the contracted spin-flavor $SU(2N_f)_c~$ symmetry,
$N_f$ being the number of light flavors, which is a symmetry of QCD in the $N_c\to \infty$ limit \cite{DM93, GS84, PY}. 
In this framework commutation relations of operators like axial currents and hadron masses are constrained by 
consistency relations. The observed baryons at $N_c=3$ are identified with the low lying spin states of an infinite representation of 
the contracted symmetry.  The second framework makes use of the spin-flavor $SU(2N_f)$ algebra, with an explicit representation 
of operators that act on a space of states constructed as tensor products of  $N_c$ valence quarks \cite{CGO94}.
Both approaches are consistent and  deliver equivalent results order by order in the $1/N_c$ expansion. From the practical point 
of view, however, the second one is easier to work with, especially at subleading orders in $1/N_c$,   and for this reason it has been 
chosen in most analyses. Another advantage in  this approach, is the possibility of using the  language of the constituent quark model, 
as applied to the spin-flavor degrees of freedom, without any loss of generality.  

The study of excited baryons is not free of difficulties.
Although a significant amount of symmetry in the form of a contracted $SU(2N_f)_c$ is always present in the  $N_c \to \infty$ limit 
\cite{PY,PS03}, there is no strict spin-flavor symmetry in that limit. 
Indeed, as it was shown in \cite{Goi97}, spin-orbit interactions  break spin-flavor symmetry at  ${\cal{O}}(N_c^0)$ in states 
belonging to mixed symmetric spin-flavor representations, and configuration mixing, {\it i.e.}, mixing of states belonging 
to different spin-flavor multiplets in general occurs at  ${\cal{O}}(N_c^0)$ as well. The use of spin-flavor symmetry as a 
zeroth order approximation is therefore not warranted for excited baryons. 
However, a phenomenological fact is that spin-orbit interactions are very small (in the real world with $N_c=3$ they have 
a magnitude expected for ${\cal{O}}(N_c^{-2})$ effects), and since all sources of ${\cal{O}}(N_c^0)$  spin-flavor breaking, including the 
configuration mixing, requires such orbital interactions,  it is justified to treat them in practice as subleading.
Thanks to this observation, the usage of spin-flavor  $SU(2N_f)$ as the zeroth order  symmetry is justified. 
A second problem is posed by the fact that excited baryons have finite widths.
The impact of this on the analyses of the masses  is  not fully clarified yet. One likely possibility is that their effects  are  
included in the effective parameters that determine the  masses'  $1/N_c$ expansion. 
This is an  issue that has  been recently considered in Ref. \cite{CL03}. 

The analysis of the $[{\mathbf{56}},2^+]$ masses is made  along the lines established in  previous investigations of
the $[\mathbf{70},1^-]$ baryons \cite{Goi97,CCGL,SGS}.
The $[{\mathbf{56}},2^+]$ multiplet contains  two $SU(3)$ octets with  total angular
momentum $J=3/2$ and  $5/2$, and four decuplets
with  $J=1/2$, $3/2$, $5/2$ and $7/2$, as listed in  Table \ref{multiplet}. Note that the octets have spin $S=1/2$ while the
decuplets have spin $S=3/2$ as in the ground state baryons. For non-strange states this
is the $I=S$ rule. 
The states are obtained by coupling the orbital part with $\ell=2$ to the
spin-flavor symmetric states, namely,
\bea
 |J\, J_z  ;  S; (p=2S,q), Y, I\, I_z \rangle_{_{\rm Sym}}
&=& \sum_{m,~S_z}
\left(
    \begin{array}{cc|c}
        S   & \ell & J   \\
        S_z & m & J_z
    \end{array}
\right)
 |S\, S_z  ; (p,q), Y,I\,I_z \rangle_{_{\rm Sym}}\mid \ell=2 \  m \rangle
\eea
where $(p,q)$ label the $SU(3)$ representation and $Y$ stands for the hypercharge.
Note that, unlike the states in mixed-symmetric representations where excited and core 
quarks have to be distinguished for the purpose of building a basis of mass operators, 
such a distinction is unnecessary for the symmetric representation.

The mass operator can  be expressed as a string of terms expanded in $1/N_c$:
\begin{equation}
H_{\rm mass}=\sum c_i O_i +\sum b_i \bar{B}_i
\end{equation}
where the operators $O_i$ are $SU(3)$ singlets and the operators $\bar{B}_i$ provide $SU(3)$ breaking 
and are defined to have vanishing matrix elements between
non-strange states.
The effective coefficients $c_i$ and $b_i$ are reduced matrix elements that encode the  QCD dynamics  
and they are determined by a fit to the empirically known masses.

 The operators $O_i$ and $\bar{B}_i$ can be expressed as
positive parity and rotationally invariant  products of generators
of $SU(6) \otimes O(3)$ as it has been explained elsewhere \cite{Goi97}.
A generic $n$-body operator has the structure
\bea
O^{(n)} &=& \frac{1}{N_c^{n-1}} \ O_{\ell} \ O_{SF} ,
\eea
where the factors $O_{\ell}$ and  $ O_{SF}$  can be expressed in terms of products of generators of the
orbital group $O(3)$ $(\ell_i)$, and of the  spin-flavor group $SU(6)$  ($S_i,~ T_a$ and $G_{ia}$), respectively.
The explicit $1/N_c$ factors originate in the $n-1$ gluon exchanges required to
give rise to an $n$-body operator.
The matrix elements of operators may also carry a nontrivial $N_c$ dependence due to coherence
effects\cite{DM93}:
for the states considered, $G_{ia}$ ($a=1,2,3$) and $T_8$  have matrix elements of  ${\cal{O}}(N_c)$,
while  the rest of the generators have matrix elements of zeroth  order.

 At each order in $1/N_c$ and $\epsilon$, where the latter parameter measures $SU(3)$ breaking, 
there is a basis of operators. The construction of these bases is straightforward, and the operators are listed in 
Table \ref{operators}. The corresponding  matrix elements between the states belonging to 
the $[\mathbf{56},2^+]$ multiplet are given in Tables \ref{singlets} and \ref{octets}. 
Note that the  operators used in the analysis of the  $[\mathbf{70},1^-]$ masses are reduced to the operators given 
here in Table \ref{operators}. This can be shown using
 reductions, valid for the symmetric representation, of matrix elements involving excited 
quark and/or core operators, such as:
\begin{eqnarray}
\langle {\rm Sym}\mid s_i \mid {\rm Sym} \rangle &=&
\frac{1}{N_c}  \langle {\rm Sym}\mid S_i \mid {\rm Sym} \rangle \\
\langle {\rm Sym}\mid S^c_i \mid {\rm Sym} \rangle &=&
\frac{N_c-1}{N_c}  \langle {\rm Sym}\mid S_i \mid {\rm Sym} \rangle,
~~~~{\rm etc.}, \nonumber
\label{newreda}
\end{eqnarray}
where $S=s+S^c$,  $s$  being the spin operator acting only on one quark (the excited one for instance), 
and $S^c$ acts on the remaining  $(N_c-1)$ core quarks. Similarly, relations for   two-body 
operators can also  be derived, {\it e.g.}:
\begin{eqnarray}
\langle {\rm Sym}\mid s_i G^c_{ja} \mid {\rm Sym} \rangle &=&
\langle {\rm Sym}\mid s_i (G_{ja}-g_{ja}) \mid {\rm Sym} \rangle \\
&=& \frac{1}{N_c}
\langle {\rm Sym}\mid  S_i G_{ja}-\frac{1}{4}\delta_{ij} T_a  - \frac{i}{2} \epsilon_{ijk} G_{ka})
\mid {\rm Sym} \rangle\nonumber
\label{newredb}
\end{eqnarray}
An important  observation is that in the present case  there is no $SU(3)$ singlet operator breaking 
spin-flavor symmetry at ${\cal{O}}(N_c^0)$. In particular, operators involving the $O(3)$ generators, that in the mixed-symmetric 
spin-flavor representations can be  ${\cal{O}}(N_c^0)$, are demoted
to ${\cal{O}}(1/N_c)$ in the spin-flavor symmetric representation.
At  ${\cal{O}}(N_c^{-1})$ only two singlet operators appear, the spin-orbit operator $O_2$ and 
the hyperfine operator $O_3$, both being two-body operators. At order $\epsilon$ there is one 
operator  ${\cal{O}}(N_c^0)$, namely $\bar{B}_1$ and two operators  ${\cal{O}}(N_c^{-1})$, namely $\bar{B}_{2,3}$.

Note that in the  $\mathbf{56}$-plet  there are no state mixings in
the $SU(3)$ symmetric limit. Only the operator $\bar{B}_2$ induces mixings. 
The  mixings affect  the octet and decuplet  $\Sigma^{(\mathbf{8}),(\mathbf{10})}$ and  
$\Xi^{(\mathbf{8}),(\mathbf{10})}$ states, and in the limit of
isospin symmetry there are four mixing angles, namely $\theta_J^{\Sigma,\Sigma'}$ and
$\theta_J^{\Xi,\Xi'}$ with $J=3/2, 5/2$. The physical states are 
given by $\Sigma_J = \Sigma^{(\mathbf{8})}_J \cos \theta_{J}^{\Sigma,\Sigma'} 
+  \Sigma^{(\mathbf{10})}_J \sin \theta_{J}^{\Sigma,\Sigma'}$
and $\Sigma'_J = -\Sigma^{(\mathbf{8})}_J \sin \theta_{J}^{\Sigma,\Sigma'} 
+  \Sigma^{(\mathbf{10})}_J \cos \theta_{J}^{\Sigma,\Sigma'} $
and in a similar way for the cascades. The mixing angles are determined by the ratio of the 
matrix elements of the operator $\bar{B}_2$ to the spin-flavor mass splitting induced by 
the  ${\cal{O}}(N_c^{-1})$ singlet operators. This implies that the mixing angles 
are ${\cal{O}}(\epsilon N_c^0)$. The mixings affect the mass eigenvalues at  ${\cal{O}}(\epsilon^2/N_c)$, 
which is beyond the accuracy of the present analysis.

Since there are twenty four  independent masses in the isospin symmetric limit,
and the basis  consists of  six operators,  there are eighteen mass relations that hold independently 
of the values of the coefficients $c_i$ and $b_i$. These relations are depicted in Table \ref{masrel}.
In addition to the  Gell-Mann Okubo (GMO) relations for each octet (two such relations) and the equal 
spacing relations
(EQS) for each decuplet (eight such relations), there are eight relations that involve states belonging 
to different $SU(3)$ multiplets as well as different values of $J$:  the first  three  in the
Table \ref{masrel} involve only the masses of non-strange states, while  the remaining five   relations
have been chosen in such a way that several of them  can be tested directly with the available data. 
These latter eight relations  provide a useful test of the validity of the $1/N_c$ expansion as implemented 
in this analysis.
The GMO and EQS
relations cannot be tested due to the scarcity of information on strange baryons.
If the one and two star states are excluded, there are four relations that can be
tested, namely, the three non-strange ones (1, 2 and 3) and  relation (4).
If the one and two star states are included (three such states), 
there are three addditional relations   that can be tested, namely (5, 6 and 7).
In all cases they  are found to be satisfied within the experimental errors.
Following \cite{JL95}, in order to compare to what extent the empirical accuracies of the 
mass relations match the theoretical expectations,  each of the mass relations in Table \ref{masrel} is cast
in the form ${\rm LHS}={\rm RHS}$ with the left hand side (LHS) and right hand side (RHS) possessing only
terms with positive coefficients.
The accuracy of the mass relations is then defined as $|{\rm LHS}-{\rm RHS}|/[({\rm LHS}+{\rm RHS})/2]$. 
These ratios are  ${\cal{O}}(\epsilon^2 N_c^{-2})$ for the GMO and EQS relations, and 
  ${\cal{O}}( N_c^{-3})$, ${\cal{O}}( \epsilon^2  N_c^{-2})$ and/or  ${\cal{O}}( \epsilon  N_c^{-3})$  for the others. 
For $N_c=3$, and  $\epsilon\sim 1/3$, 
 the  ratios associated with  the relations  (1) to (8) in Table \ref{masrel} 
are estimated  to be  of the order of  $4 \%$.
The  ratios  obtained with the physical masses are listed in the last
column of Table \ref{masrel}, and  they are  within  that estimated theoretical range. 
It is important to
emphasize that all these empirically verified  relations  represent a genuine test of spin-flavor symmetry and
its breaking according to the $1/N_c$ expansion, as pointed out above. The fact that they are all 
verified to the expected accuracy is remarkable  and gives strong support to the analysis based on the premises of this work.

The fit to the available data, where states with three or more stars in the Particle
Data Listings are included, leads to the effective constants $c_i$ and $b_i$ shown in Table \ref{operators} and the
results for the masses shown in Table \ref{multiplet}, where fourteen of them are predictions.
 The $\chi^2_{\rm dof}$ of the fit is 0.7, where the number of degrees of freedom ({\rm dof}) is equal to four. 
The errors shown for the predictions in Table \ref{multiplet} are obtained propagating the errors of 
the coefficients in Table \ref{operators}. There is also a systematic error  ${\cal{O}}( N_c^{-2})$, 
resulting from having included only  operators up to  ${\cal{O}}( N_c^{-1})$ in the analysis,
which can be roughly estimated to be around 30~MeV. 

In Table \ref{multiplet}  the partial contributions from each operator to the mass of the
different members of the multiplet are also shown.
The operator $O_1$ provides the spin-flavor singlet mass of
about 1625 MeV. The  breaking of spin-flavor symmetry by the $SU(3)$ singlet operators
is essentially given in its entirety by the hyperfine interaction $O_3$, that produces a splitting 
between octet and decuplet states of approximately $240$ MeV, while the
spin-orbit operator $O_2$ is rather irrelevant inducing spin-flavor breaking mass
shifts of less than 30 MeV. Note that $O_2$ is the sole source of the splittings between the two N 
states and also between the $\Delta$ states. The weakness of $O_2$ is thus very convincingly established.

The breaking of $SU(3)$ is dominated by the operator $\bar{B}_1$, which gives a
shift of about 200~MeV per unit of strangeness. The main role of the subleading
operators $\bar{B}_{2,3}$ is to provide the  observed $\Lambda-\Sigma$ splittings
in the octets, and the different splittings between the $N$ and the average
$\Lambda-\Sigma$ masses in the two octets, and the $\Sigma-\Delta$ splitting
in the $J=7/2$ decuplet.  Finally, $\bar{B}_{2}$ gives the only contributions to the state mixings.
The  mixing angles  that result from the fit are: $\theta^{\Sigma,\Sigma'}_{3/2} = -0.16$,
$\theta^{\Sigma,\Sigma'}_{5/2} = -0.26$, $\theta^{\Xi,\Xi'}_{3/2} = -0.21$ and
$\theta^{\Xi,\Xi'}_{5/2} = -0.19$ (in radians). 

 The  better established $\Lambda-\Sigma$ splitting in the $J=5/2$ octet is
almost 100 MeV, while the other splitting in the $J=1/2$ octet is small and
slightly negative. The latter one involves however the one star state
 $\Sigma(1840)$, which might  also be assigned to the radially excited
$\mathbf{56'}$.  The large $N_c$ analysis implies that these splittings are
${\cal{O}}(\epsilon/N_c)$, and are produced only by the operators $\bar{B}_2$
and  $\bar{B}_3$. The result from the fit indicates that the
$\Lambda_{5/2}-\Sigma_{5/2}$ receives a contribution of 63 MeV from $\bar{B}_3$
and 40 MeV from $\bar{B}_2$. 
It is interesting to observe 
that several mass splitting differences receive only contributions from $\bar{B}_2$ as it is obvious from Table \ref{octets}. 
These involve the splittings in the octets
$(\Lambda_{5/2}-N_{5/2})-(\Lambda_{3/2}-N_{3/2})$, $(\Sigma_{5/2}-N_{5/2})-(\Sigma_{3/2}-N_{3/2})$ and
$(\Xi_{5/2}-N_{5/2})-(\Xi_{3/2}-N_{3/2})$, and the decuplet splittings $(\Sigma_{J}-\Delta_{J})-(\Sigma_{J'}-\Delta_{J'})$,
$(\Xi_{J}-\Delta_{J})-(\Xi_{J'}-\Delta_{J'})$ and $(\Omega_{J}-\Delta_{J})-(\Omega_{J'}-\Delta_{J'})$. 
Further information on  these splittings would allow to pin 
down with better confidence the relevance of $\bar{B}_2$. The  fit implies for instance that the  
contribution of $\bar{B}_2$ to the $\Omega_{1/2}-\Omega_{7/2}$ splitting is about  $225\pm 100$ MeV, 
a rather large effect. The operator $\bar{B}_2$  involves the orbital angular momentum operator, and since in all
other known cases where orbital couplings occur their  effects are
suppressed, the same would be expected here. The naive expectation is that the coefficient 
of  $\bar{B}_2$ would be of order  $2\sqrt{3}\; \epsilon$  times the coefficient of $O_2$. 
It is in fact substantially larger. However,  this result is  not very conclusive, because $b_2$ is 
largely determined by only a few inputs 
resulting in a rather large relative error for this parameter. Related to this, the sign of the
coefficient of $\bar B_2$ determines the ascending or descending ordering of the masses of strange states 
in the decuplet  as $J$ increases. In the present analysis the higher $J$ states are lighter. However, 
the structure of $SU(3)$ breaking splittings cannot be established better because of 
the rather small number of strange states available for the fit. This is perhaps the most important motivation 
for further experimental and lattice QCD study of the still non-observed  states.

It is of interest to draw some comparisons among the analyses carried out in
previous works, that include the ground state baryons \cite{Jenk1,JL95},
the $[\mathbf{70},1^-]$ baryons \cite{CCGL,SGS}, the Roper multiplet  $[\mathbf{56'},0^+]$
\cite{CC00}, and the present analysis. At the level of $SU(3)$ singlet operators
the hyperfine interaction is ${\cal{O}}(1/N_c)$ in all cases. It is interesting to compare the strength of 
the hyperfine interaction in the different multiplets estimating the strength of the quark pairwise hyperfine interaction. In the large $N_c$ limit that strength should be the same for different low lying excited states.  For the ground state baryons the hyperfine 
operator, up to terms proportional to the identity operator, is given by: $\sum_{i \neq j} s_{(i)} \cdot s_{(j)}$ 
where the indices $i$, $j$ run from 1 to $N_c$. The ground state $\Delta-N$ splitting then 
gives a strength of about 100 MeV for this operator. In excited  states with  $\ell=1$, 
the  results from the $\mathbf{70}$-plet depend in general on the mixing angles used as an input \cite{PS03}.
For the particular choice of the angles used in the analyses \cite{CCGL, CaCa98,SGS} the hyperfine interaction involving the excited 
quark and quarks in the core (the operator  $O_7$
in \cite{SGS})  is suppressed, indicating that the hyperfine interaction is predominantly short range.
The relevant hyperfine interaction is in this case  the one involving the $N_c-1$ quarks in the 
core, {\it i.e.}, the indices $i$ and $j$ run only over the quarks in the core. This leads in 
the $[\mathbf{70},1^-]$   to a strength of about 160 MeV.  In  the $[\mathbf{56},2^+]$,  a reasonable assumption is that a single quark is excited with $\ell=2$.  From the result obtained in the $[\mathbf{70},1^-]$, it is expected that the excited quark will also  have negligible participation in the hyperfine interaction. This  will be given essentially by the hyperfine interaction of the core quarks.  Using the relation  $ \left(1 - \frac{2}{N_c}\right) S^2 = (S^{c})^2 - \frac{3}{4}\mathbbm{1} $ 
valid in states belonging to the symmetric spin-flavor representation, the result of the fit implies a  strength of  240 MeV.  In the Roper $[\mathbf{56'},0^+]$ multiplet 
the situation is less clear, as all quarks may participate of the hyperfine interaction. The average 
strength in the core turns out to be  about 160 MeV. These results  indicate  an increase in the  
strength of the hyperfine
interaction in going from the ground state baryons to excited baryons. This suggests the presence of 
an underlying  dynamical mechanism that 
might  be possible to identify in  specific models. The other $SU(3)$ singlet interaction 
common to all multiplets,  the spin-orbit interaction, is  weak in the two known cases, 
namely  $[\mathbf{70},1^-]$ and $[\mathbf{56},2^+]$.

 The $SU(3)$ breaking operator ${\bar B_1}=-{\cal S}$ gives a mass shift per unit
of strangeness
of about 200 MeV in all multiplets considered, which is in line with the
  the value
of the strange quark mass. The operator $l_i g_{i8}$ which contributes at ${\cal{O}}(\epsilon N_c^0)$
in the $\mathbf{56}$-plet and at ${\cal{O}}(\epsilon / N_c)$ in the $\mathbf{70}$-plet, 
carries  coefficients of similar size but different sign. This issue can be further clarified 
when the role of $\bar{B}_2$ is better established.

\begin{acknowledgements}
We thank Winston Roberts for useful comments on the manuscript. JLG thanks the Physics Department of TANDAR (Argentina) for the kind hospitality.
This work was partially supported by the National Science Foundation (USA) through grant
\#~PHY-9733343, by the  Department of Energy through  contract DOE-AC05-84ER40150 (JLG) and
DOE-FG02-96ER40945 (CS), and by ANPCYT (Argentina) grant PICT00-03-0858003(NNS). The authors acknowledge support from Fundaci\'on Antorchas 
(Argentina).
NNS would  also  like to acknowledge support by ICTP (Trieste, Italy) during his stay at the ECT$^*$ (Trento, Italy). 
\end{acknowledgements}

\pagebreak

\begin{table}[tbp]
\vspace{2cm}
\begin{tabular}{llrrl}
\hline
\hline
Operator & \multicolumn{4}{c}{Fitted coef. (MeV)}\\
\hline
\hline
$O_1 = N_c \ {\mathbbm{1}} $                                    & \ \ \ $c_1 =  $  & 541 & $\pm$ & \   4   $\ $ \\
$O_2 =\frac{1}{N_c} l_i \ S_i$                                  & \ \ \ $c_2 =  $  &  18 & $\pm$ & 16   $\ $ \\
$O_3 = \frac{1}{N_c}S_i S_i$                                    & \ \ \ $c_3 =  $  & 241 & $\pm$ & 14   $\ $ \\
\hline
$\bar B_1 = -{\cal S} $                                         & \ \ \ $ b_1 = $  & 206 & $\pm$ & 18   $\ $ \\
$\bar B_2 = \frac{1}{N_c} l_i G_{i8}-\frac{1}{2 \sqrt{3}} O_2$  & \ \ \ $ b_2 = $  & 104 & $\pm$ & 64   $\ $ \\
$\bar B_3 = \frac{1}{N_c} S_i G_{i8}-\frac{1}{2 \sqrt{3}} O_3$ \ \ \ \ \  & \ \ \ $ b_3 = $  & 223 & $\pm$ & 68   $\ $ \\
\hline \hline
\end{tabular}
\caption{List of operators and the coefficients resulting from the fit with $\chi^2_{\rm dof} =0.7 $  }
\label{operators}
\vspace{4cm}
\end{table}

\begin{table}
\[
\begin{array}{crrr}
\hline
\hline
           & \ \ \ \ \ \ \ \  O_1  & \ \ \ \ \ \ \ O_2  & \ \ \ \ \ \ \ O_3  \\
\hline
^28_{3/2}  & N_c  & - \fr{3}{2 N_c} & \fr{3}{4 N_c}  \\
^28_{5/2}  & N_c  &   \fr{1}{  N_c} & \fr{3}{4 N_c}  \\
^410_{1/2} & N_c  & - \fr{9}{2 N_c} & \fr{15}{4 N_c} \\
^410_{3/2} & N_c  & - \fr{3}{  N_c} & \fr{15}{4 N_c} \\
^410_{5/2} & N_c  & - \fr{1}{2 N_c} & \fr{15}{4 N_c} \\
^410_{7/2} & N_c  &   \fr{3}{  N_c} & \fr{15}{4 N_c} \\[0.5ex]
\hline
\hline
\end{array}
\]
\caption{Matrix elements of  $SU(3)$ singlet operators.}
\label{singlets}
\end{table}

\clearpage

\begin{table}
\[
\begin{array}{cccc}
\hline
\hline
  &  \hspace{ .6 cm}  {\bar B}_1 \hspace{ .6 cm}  &
     \hspace{ .6 cm}  {\bar B}_2 \hspace{ .6 cm}  &
     \hspace{ .6 cm}  {\bar B}_3 \hspace{ .6 cm}   \\
\hline
N_{J}       & 0 &                      0  &                           0  \\
\Lambda_{J} & 1 &   \fr{ 3 \sqrt{3}\ a_J}{4 N_c} &   - \fr{3 \sqrt{3}}{8 N_c}   \\
\Sigma_{J}  & 1 & - \fr{  \sqrt{3}\ a_J}{4 N_c} &     \fr{  \sqrt{3}}{8 N_c}   \\
\Xi_{J}     & 2 &   \fr{  \sqrt{3}\ a_J}{N_c}   &   - \fr{  \sqrt{3}}{2 N_c}   \\ [0.5ex]
\hline
\Delta_{J}  & 0 &                      0  &                           0  \\
\Sigma_{J}  & 1 &  \fr{3 \sqrt{3}\ b_J}{4 N_c} &   - \fr{ 5 \sqrt{3}}{8 N_c}  \\
\Xi_{J}     & 2 &  \fr{3 \sqrt{3}\ b_J}{2 N_c} &   - \fr{ 5 \sqrt{3}}{4 N_c}  \\
\Omega_{J}  & 3 &  \fr{9 \sqrt{3}\ b_J}{4 N_c} &   - \fr{15 \sqrt{3}}{8 N_c}  \\ [0.5ex]
\hline
\Sigma_{3/2} - \Sigma'_{3/2}    &  0 &  \fr{  \sqrt{3}}{2 N_c}   &   0   \\
\Sigma_{5/2} - \Sigma'_{5/2}    &  0 &  \fr{  \sqrt{3}}{2 N_c}   &   0   \\
\Xi_{3/2} - \Xi'_{3/2}          &  0 &  \fr{ \sqrt{42}}{6 N_c}   &   0   \\
\Xi_{5/2} - \Xi'_{5/2}          &  0 &  \fr{ \sqrt{42}}{6 N_c}   &   0   \\[0.5ex]
\hline
\hline
\end{array}
\]
\caption{Matrix elements of $SU(3)$ breaking operators. Here, $a_J = 1,-2/3$ for $J=3/2, 5/2$, respectively and
$b_J = 1, 2/3, 1/9, -2/3$ for $J=1/2, 3/2, 5/2, 7/2$, respectively.}
\label{octets}
\end{table}

\begin{table}
\[
\begin{array}{crclc}
\hline
\hline
& & & & Accuracy \\
(1) & \Delta_{5/2} - \Delta_{3/2}     & = & N_{5/2} - N_{3/2}  & 0.6 \% \\
(2) & 5 (\Delta_{7/2} - \Delta_{5/2}) & = & 7 (N_{5/2} - N_{3/2}) & 1.8 \% \\
(3) & \Delta_{7/2} - \Delta_{1/2} & = & 3 (N_{5/2} - N_{3/2}) & 1.5 \% \\
\hline
(4) & 8 (\Lambda_{3/2} - N_{3/2}) + 22 ( \Lambda_{5/2} - N_{5/2})
&=&   15 (\Sigma_{5/2}- \Lambda_{5/2}) + 30 (\Sigma_{7/2}-\Delta_{7/2}) & 0.4 \%  \\
(5) & \Lambda_{5/2} - \Lambda_{3/2} + 3(\Sigma_{5/2} - \Sigma_{3/2}) & = &  4 (N_{5/2} - N_{3/2}) & 1.7 \% \\
(6) & \Lambda_{5/2} - \Lambda_{3/2} + \Sigma_{5/2} - \Sigma_{3/2} & = &  2 (\Sigma'_{5/2} - \Sigma'_{3/2}) &  0.5 \%  \\
(7) & 7 \; \Sigma'_{3/2} + 5 \; \Sigma_{7/2} &=& 12 \; \Sigma'_{5/2}  & 0.5 \%  \\
(8) & 4 \; \Sigma_{1/2} + \Sigma_{7/2} &=& 5 \; \Sigma'_{3/2} & \\
\hline
{\rm (GMO)  }    &   2 (N + \Xi) &=& 3 \; \Lambda + \Sigma  \\
{\rm (EQS)  }       &  \Sigma - \Delta &=& \Xi - \Sigma = \Omega - \Xi  \\
\hline
\hline
\end{array}
\]
\caption{The 18 independent mass relations include
the  GMO relations for the two octets and the two EQS for each of the four decuplets. The accuracy is calculated
as explained in the text.}
\label{masrel}
\end{table}

\pagebreak

\begin{table}
\begin{tabular}{crrrrrrcccccc}\hline \hline
                    &      \multicolumn{7}{c}{1/$N_c$ expansion results}        &    &                   \\
\cline{2-8}
                    &      \multicolumn{6}{c}{Partial results} & \hspace{.2cm} Total  \hspace{.2cm}  &
                                              \hspace{.2cm}   Empirical   \hspace{.2cm}   \\
\cline{2-7}
                    &      \hspace{.5cm} $O_1$  & \hspace{.4cm}  $O_2$ &
                            \hspace{.4cm} $O_3$  & \hspace{.4cm} $\bar B_1$  &
                           \hspace{.4cm} $\bar B_2$  & \hspace{.4cm} $\bar B_3$ 
                                                                                                & \hspace{3cm}  &   \\
\hline
$N_{3/2}$           & $1623$  &$  -9 $   & $60 $ &  0     &   0    &   0     &  $ 1674\pm15  $    &  $ 1700\pm50 $ \\
$\Lambda_{3/2}$     &          &          &      &$206$&$ 45$ & $-48$&$   1876\pm39 $    &  $ 1880\pm 30 $ \\
$\Sigma_{3/2}$      &          &          &      &$206$&$-15$ &$  16$ &$   1881\pm25$     &  $   (1840)   $ \\
$\Xi_{3/2}$         &          &          &      &$412$&$ 60$ &$ -64$& $  2081\pm57 $    &                 \\
\hline
$N_{5/2}  $         &  $1623$  & $ 6    $ & $60$ &  0     &   0    &   0     & $  1689\pm14  $   &  $ 1683\pm 8  $ \\
$\Lambda_{5/2}$     &	       &          &      &$206$&$-30$&$ -48$& $  1816\pm33$     &  $ 1820\pm 5  $ \\
$\Sigma_{5/2}$      &          &          &      & $206$       &$  10$&$  16 $& $  1920\pm24 $    &  $ 1918\pm 18 $ \\
$\Xi_{5/2}$         &          &          &      &$412$&$-40$&$ -64$&$   1997\pm49 $    &                 \\
\hline
$\Delta_{1/2}$      &  $1623$  & $ -27$   & $301$ &0     &   0    &   0     &  $ 1897\pm32 $    &  $ 1895\pm 25 $ \\
$\Sigma^{}_{1/2}$   &          &          &       &$206$&$ 45$&$ -80$&$   2068\pm52 $    &                 \\
$\Xi^{}_{1/2}$      &          &          &       &$412$&$ 90$&$-161$&$   2237\pm88  $   &                 \\
$\Omega_{1/2}$      &          &          &       &$618$&$135$&$-241$&$   2408\pm127$    &                 \\
\hline
$\Delta_{3/2}$      & $1623$   & $-18$    & $301$    &  0     &   0    &    0    & $  1906\pm27 $   & $  1935\pm 35$  \\
$\Sigma'_{3/2}$     &          &          &          &$206$&$ 30$&$-80 $& $  2061\pm44  $  & $ (2080)     $  \\
$\Xi'_{3/2}$        &          &          &          &$412$&$ 60$&$-161$& $  2216\pm76 $   &                 \\
$\Omega_{3/2}$      &          &          &          &$618$&$ 90$&$-241$& $  2373\pm110 $  &                 \\
\hline
$\Delta_{5/2}$      & $1623$   &  $  -3 $ &  $301$ &  0     &   0    &    0    & $  1921\pm21 $  & $ 1895 \pm 25 $ \\
$\Sigma'_{5/2}$     &          &          &        &$206$&$   5$&$-80$ &  $ 2051\pm37 $  & $ (2070)      $ \\
$\Xi'_{5/2}$        &          &          &        &$412$&$  10$&$-161$& $  2181\pm64  $ &                 \\
$\Omega_{5/2}$      &          &          &        &$618$&$  15$&$-241$&$   2313\pm94 $  &                 \\
\hline
$\Delta_{7/2}$      &  $1623$  & $  18$   & $301$ &  0     &   0    &    0    &  $ 1942\pm27 $  & $ 1950\pm 10 $  \\
$\Sigma^{}_{7/2}$   &          &          &       &$206$&$-30$&$-80 $&$   2036\pm44 $  & $ 2033\pm 8   $ \\
$\Xi^{}_{7/2}$      &          &          &       &$412$&$-60$&$-161$&$   2131\pm76 $  &                 \\
$\Omega_{7/2}$      &          &          &       &$618$&$-90$&$-241$&$   2229\pm110 $ &                 \\
\hline \hline
\end{tabular}
\caption{Masses (in MeV) predicted by the $1/N_c$ expansion as compared with the 
empirically known masses. The partial contributions to each mass by the operators 
in the basis are shown. Those partial contributions in blank are equal to the one 
above in the same column.}
\label{multiplet}
\end{table}

\end{document}